\documentclass[reprint,superscriptaddress,showpacs,amsmath,amssymb,aps,prx]{revtex4-2}

\usepackage{mathtools}
\usepackage{natbib}
\usepackage{siunitx}
\usepackage{array}
\usepackage{xcolor}
\usepackage{subfigure} 
\usepackage{amsmath} 
\usepackage{amsfonts}
\usepackage[colorlinks=true,allcolors=blue]{hyperref}
\usepackage{amssymb}
\usepackage{graphicx}
\usepackage{hyperref}

\providecommand{\vect}[1]{{\boldsymbol{#1}}}

\begin{document}

\title{
Nonzero Skyrmion Hall effect in topologically trivial structures}

\author{Robin Msiska}
\affiliation{Faculty of Physics, University of Duisburg-Essen, 47057 Duisburg, Germany}
\author{Davi R. Rodrigues}
\affiliation{Department of Electrical and Information Engineering, Politecnico di Bari, 70126 Bari, Italy}
\author{Jonathan Leliaert}
\affiliation{Department of Solid State Sciences, Ghent University, 9000 Ghent, Belgium}
\author{Karin Everschor-Sitte}
\affiliation{Faculty of Physics, University of Duisburg-Essen, 47057 Duisburg, Germany}
\affiliation{Center for Nanointegration Duisburg-Essen (CENIDE), 47057 Duisburg, Germany}

\date{\today}

\begin{abstract}
It is widely believed that the skyrmion Hall effect, often disruptive for device applications, vanishes for overall topologically trivial structures such as (synthetic) antiferromagnetic skyrmions and skyrmioniums due to a compensation of Magnus forces. 
In this manuscript, however, we report that in contrast to the case of spin-transfer torque driven skyrmion motion, this notion
is generally false for spin-orbit torque driven objects. We show that the skyrmion Hall angle is directly related to their helicity and imposes an unexpected roadblock for developing faster and lower input racetrack memories based on spin-orbit torques. 
\end{abstract}

\maketitle

%%%%%%%%%%%%%%%%%%%%%%%%%%%%%%%%%%%%%%%%%%%%%%%%%%%%
\section{Introduction}
\label{sec:Intro}
%%%%%%%%%%%%%%%%%%%%%%%%%%%%%%%%%%%%%%%%%%%%%%%%%%%%

Magnetic skyrmions are localized whirl-like magnetic textures with a non-trivial topology~\cite{Back2020}. Skyrmions driven by electric currents have been shown to exhibit a significant transverse component in addition to their longitudinal current-induced motion along a track. This deviation in skyrmion motion has been termed the skyrmion Hall effect~\cite{Everschor2011, Jiang2016, Litzius2017}. While the physics of the skyrmion Hall effect is fascinating, and allows skyrmions to evade defects~\cite{Sampaio2013, Iwasaki2013, Rosch2013, Mueller2015}, it often imposes a challenge for skyrmion-based devices~\cite{Kang2016a, Fert2017, Everschor-Sitte2018, Back2020, Zhang2020}. In particular for skyrmion race track proposals~\cite{Fert2013, Tomasello2014, Zhang2015c}, the driving speed of magnetic skyrmions is limited by the skyrmion Hall effect, as beyond a certain drive they get pushed into the boundary of the sample.

Numerous suggestions have been made to suppress or eliminate the skyrmion Hall effect~\cite{Huang2017a, Kim2018, Gobel2019a, Gobel2019, Zhang2020d, Zarzuela2020}. Prominent among them is the idea of using combined skyrmion structures with opposite winding numbers such that the composite structure is topologically trivial. Among them there are skyrmion structures in (synthetic) antiferromagnetic materials, which have the additional advantages of obeying faster dynamics as well as small stray fields, and being insensitive to external fields~\cite{Baltz2018, Gomonay2018}. Another proposal is the use of skyrmioniums~\cite{Zhang2018,Zhang2016,Kolesnikov2018,Gobel2019}. 
In these systems there is the wide belief that the opposite topological charges of the two substructures lead to a cancellation of the acting Magnus forces~\cite{Zhang2015j, Zhang2015b, Zhang2016, Barker2016, Gobel2017,Gomonay2018, Kolesnikov2018,Gobel2019}. 

\begin{figure}[t]
  \centering
  \includegraphics[width=1.\linewidth]{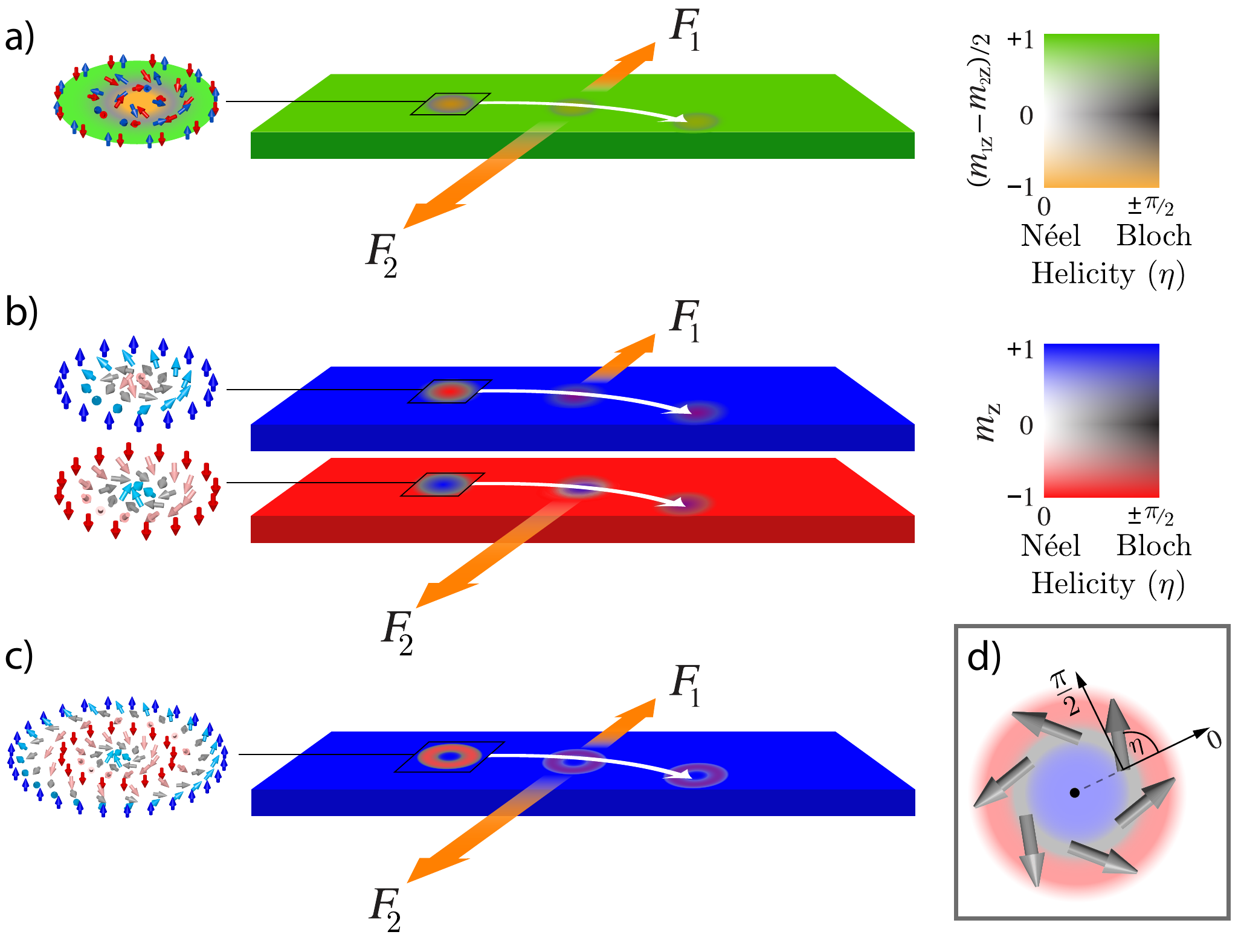}  
\caption{Net Skyrmion Hall effect in topologically trivial skyrmionic structures. The Magnus forces acting on a) the different sublattices for a antiferromagnetic skyrmion, b) the different layers for a synthetic antiferromagnetic skyrmion, and  c) the different co-centric skyrmions for the skyrmionium do not compensate each other. The grey scale encodes the helicity and the color shows the out-of plane component of a) the Néel order parameter; b) and c) the magnetization. 
In d) we show the definition of the helicity as the azimuthal angle of the magnetization in a skyrmionic structure. Notice that $\eta =0$ corresponds to an outward pointing N\'eel skyrmion, whereas $\eta = \pi/2$ represents a Bloch skyrmion as shown.}
\label{fig:Forces}
\end{figure}

In this work, we show, however, that this picture is generally not correct for spin-orbit-torque (SOT) driven skyrmions in (synthetic) antiferromagnets and skyrmioniums. This effect occurs as the Magnus forces acting on the different skyrmionic structures do not cancel, see Fig.~\ref{fig:Forces}. By computing the Hall angle, we reveal that there is typically a non-zero skyrmion Hall effect originating in the structure's helicity, i.e.\ the azimuthal angle of the skyrmion-like structures, see Fig.~\ref{fig:Forces} d). The helicity of a skyrmionic structure is typically determined by the twisting interactions, such as Dzyaloshinskii-Moriya interactions (DMI) and dipolar fields~\cite{Knoester2014,Kim2018, McKeever2018} and can be measured via resonant elastic x-ray scattering techniques~\cite{Zhang2018x}. 

In this manuscript we first present the SOT-driven magnetization dynamics. We derive within a Thiele approach the helicity dependence of the skyrmion Hall angle for the topologically trivial structures, i.e. the skyrmionium and the (synthetic) antiferromagnetic skyrmion. 
Our analytical results, which we confirm by means of micromagnetic simulations, contradict the usual notion that a non-zero skyrmion Hall angle must be associated to a topologically non-trivial magnetic structure.

%%%%%%%%%%%%%%%%%%%%%%%%%%%%%%%%%%%%%%%%%%%%%%%%%%%%
\section{Spin-orbit torque driven magnetization dynamics}
\label{sec:dynamics}
%%%%%%%%%%%%%%%%%%%%%%%%%%%%%%%%%%%%%%%%%%%%%%%%%%%%
The current-driven magnetization dynamics of a ferromagnetic material is well described by the Landau-Lifshitz-Gilbert equation (LLG)
\begin{equation}\label{eq:LLG}
\partial_t\vect{m}=-\gamma\vect{m}\times\vect{H}^{\rm eff}+\alpha\vect{m}\times\partial_t\vect{m}+\vect{T}(\vect{m}),
\end{equation}
where $\gamma$ is the gyromagnetic ratio and $\alpha$ is the Gilbert damping parameter. The effective magnetic field is given by $\vect{H}^{\rm eff} = -(1/M_s)\delta E[\vect{m}]/\delta \vect{m}$, where $M_{s}$ is the magnetization saturation and $E$ is the total free energy of the system. 
The term $\vect{T}(\vect{m})$ represents torques which are acting on the system. In the case of SOTs it takes the form~\cite{Slonczewski2002,Garate2010b,Hayashi2014}
\begin{equation}
\vect{T}^\text{SOT}(\vect{m})=\xi\vect{m}\times(\hat{\vect{z}}\times \vect{v}^\text{eff})+ \vect{m}\times[\vect{m}\times (\hat{\vect{z}}\times \vect{v}^\text{eff})],
\end{equation}
where $\xi$ is the field to damping like torques ratio, $\vect{v}^\text{eff} = \gamma \hbar \theta_{\textrm{Hall}} \vect{j} /(2M_{s}^2 e l)$ is the effective spin velocity with $\hbar$ being the Planck constant, 
$\theta_{\textrm{Hall}}$ is the spin Hall ratio, $\vect{j}$ is the applied current density, $e$ is the electronic charge, and $l$ the thickness of the sample~\cite{Litzius2017,Tomasello2014}.

%%%%%%%%%%%%%%%%%%%%%%%%%%%%%%%%%%%%%%%%%%%%%%%%%%%%
\section{Skyrmion Hall angle in topologically trivial structures}
\label{sec:dynamics}
%%%%%%%%%%%%%%%%%%%%%%%%%%%%%%%%%%%%%%%%%%%%%%%%%%%%

Assuming that the applied torques are weak compared to the magnetic interactions, we consider an ansatz of a rigid magnetic texture moving with drift velocity $\vect v^d$ as $\vect{m} = \vect{m}(\vect{r}- \vect{v}^d t)$. The skyrmionium and antiferromagnetic skyrmions can be described in terms of coupled skyrmions, where the index $i=1,2$ labels the corresponding skyrmion in the following. Particularly, we can describe a skyrmionium as two concentric skyrmions with different radii $R_1$ and $R_2$, and antiferromagnetic skyrmions as a pair of skyrmions belonging to two different layers/sublattices with same radius $R_1=R_2$. The current-driven dynamics of these objects can be described by the coupled Thiele equations for each skyrmion structure described by $\vect{m}_i(\vect{r}- \vect{v}^d t)$~\cite{Thiele1973} (for details, see App.~\ref{app:Thiele})

\begin{equation}\label{eq:Thiele}
	- \vect{G}_i \times  \vect{v}^d - \alpha \vect{D}_{i} \vect{v}^d + \gamma\vect{F}^{\rm int}_{i} + (\xi \, \boldsymbol{T}_{{\rm FL}\,i} + \boldsymbol{T}_{{\rm DL}\, i})\vect{v}^{\mathrm{eff}}= \vect{0}.
\end{equation}
Here $\vect{G}_i = -4 \pi Q_i \vect{e}_z$ is the gyrocoupling vector with the topological magnetic charge
\begin{equation}
	Q_{i} = \frac{1}{4\pi}\int dx\, dy\, \vect{m}_{i}\cdot\left(\partial_{x}\vect{m}_{i}\times\partial_{y}\vect{m}_{i}\right).
	\label{eq:Q}
\end{equation}
Furthermore, 
\begin{equation}
	\label{eq:Dtensor}
	(\vect{D}_{i})_{ab} = \int dx\,dy (\partial_{a}\vect{m}_{i}\cdot\partial_{b}\vect{m}_{i})
\end{equation}
is the dissipative tensor.
The force $\vect{F}^{\rm int}_{i}$ captures the interaction between the two skyrmions. In the case of skyrmioniums it is mostly due to the exchange coupling between the skyrmions while for the antiferromagnetic skyrmion it is due to the antiferromagnetic exchange between the two layers/sublattices. We notice that the $\vect{F}^{\rm int}_{1} = -\vect{F}^{\rm int}_{2}$, since the net force acting on the coupled skyrmions due to the mutual interaction vanishes.

The tensors $\boldsymbol{T}_{\mathrm{FL}}$ and $\boldsymbol{T}_{\mathrm{DL}}$ represent the field-like and damping-like spin torques.
For SOTs they are given as
\begin{subequations}\label{eq:ThieleSOTs}
	\begin{align}
		(\boldsymbol{T}^\text{SOT}_{\mathrm{FL}\,i}\vect{v}^{\mathrm{eff}})_{a} =& \int dx\,dy\, (\hat{\vect{z}}\times\vect{v}^{\mathrm{eff}})\cdot\partial_{a}\vect{m}_{i},\\
		(\boldsymbol{T}^\text{SOT}_{\mathrm{DL}\,i}\vect{v}^{\mathrm{eff}})_{a} =& \int dx\,dy\, (\vect{m}_{i}\times(\hat{\vect{z}}\times\vect{v}^{\mathrm{eff}}))\cdot\partial_{a}\vect{m}_{i}.
	\end{align}	
\end{subequations}
Here $\xi$ is the ratio between the field and damping-like torque strengths.

For the skyrmionium as well as the (synthetic) antiferromagnetic skyrmion we obtain $Q_{1} = -Q_{2}$, $\vect{D}_{1}/\vect{D}_{2}\approx f_D(R_1/R_2)$, 
$ (\boldsymbol{T}^\text{SOT}_{\mathrm{FL}\,1})/(\boldsymbol{T}^\text{SOT}_{\mathrm{FL}\,2})\approx - f_{\mathrm{FL}}(R_1/R_{2})$, 
$ (\boldsymbol{T}^\text{SOT}_{\mathrm{DL}\,1})/(\boldsymbol{T}^\text{SOT}_{\mathrm{DL}\,2})\approx f_{\mathrm{DL}} (R_1/{R_{2}})$, with positive functions $f_D, f_{\mathrm{DL}}$, and $f_{\mathrm{FL}}$ that fulfill $f_D(1)=f_{\mathrm{FL}}(1)=f_{\mathrm{DL}}(1) =1$.
Thus, the net motion of the coupled skyrmions is given by the sum of the Thiele equations for each skyrmion and simplifies approximately to
\begin{align}\label{eq:Thielesum}
 &\alpha \left[1 + f_D\left(\frac{R_{2}}{R_{1}}\right)\right]\vect{D}_{1} \vect{v}^d + \left[1 + f_\mathrm{DL} \left(\frac{R_{2}}{R_{1}}\right)\right] \boldsymbol{T}^\text{SOT}_{{\rm DL\, 1}}\vect{v}^{\mathrm{eff}}\notag\\ 
& \quad \quad + \xi \left[1 -  f_{\mathrm{FL}}\left(\frac{R_{2}}{R_{1}}\right)\right]\boldsymbol{T}^\text{SOT}_{{\rm FL\, 1}}\vect{v}^{\mathrm{eff}}= \vect{0}.
\end{align}
For a radially symmetric skyrmionic-structure $\vect{D}$ is diagonal and independent of the helicity with $(\vect{D}_{1})_{xx} = (\vect{D}_{1})_{yy} \equiv D $. The spin torque tensors $\boldsymbol{T}^\text{SOT}_{{\rm FL}}$ and $\boldsymbol{T}^\text{SOT}_{{\rm DL}}$, however, have helicity-dependent off-diagonal components,
\begin{subequations}\label{eq:SOTansatz}
	\begin{align}
		(\boldsymbol{T}_{\rm FL\,1}^{\rm SOT})_{ab} =& \tau_{\rm FL}\left( -\cos\eta \epsilon_{zab} + \sin\eta\delta_{ab}\right),\\
		(\boldsymbol{T}_{\rm DL\,1}^{\rm SOT})_{ab} =& \tau_{\rm DL} \left( \sin\eta \epsilon_{zab} + \cos\eta\delta_{ab}\right).
	\end{align}
\end{subequations}
For example, for a N\'eel skyrmion, $\eta=0 $ (Bloch skyrmion, $\eta =\pi/2$) the damping-like torques are aligned only (parallel) perpendicular to the effective spin velocity. 
The constants $D$ and $\tau_{\rm DL}$ are determined by the specific radial profile of the skyrmion-like structure.

By leveraging all contributions in Eq.~\eqref{eq:Thielesum} we derive the drift velocity $\vect{v}^d$ for the SOT-driven topologically trivial skyrmionic structures as
\begin{equation}\label{eq:SkyrmionVelocity}
	\vect{v}^{d} = \frac{1}{\alpha D} \left[1 + f_D\left(\frac{R_{2}}{R_{1}}\right)\right]^{-1}
	\left( v_{\parallel} \vect{v}^{\mathrm{eff}} + v_{\perp} (\hat{\vect{z}}\times\vect{v}^{\mathrm{eff}})\right),
\end{equation}
with the parallel and perpendicular components being
\begin{subequations}
\begin{align}
v_{\parallel}  = & \tau_{\mathrm{FL}}\, \xi \left[1 -  f_{\mathrm{FL}}\left(\frac{R_{2}}{R_{1}}\right)\right]\sin\eta \\ \notag
&+ \tau_{\mathrm{DL}}\left[1 +  f_{\mathrm{DL}}\left(\frac{R_{2}}{R_{1}}\right)\right]\cos\eta, \\
v_{\perp} = & \tau_{\mathrm{FL}} \, \xi \left[1 -  f_{\mathrm{FL}}\left(\frac{R_{2}}{R_{1}}\right)\right]\cos\eta \\ \notag
&- \tau_{\mathrm{DL}}\left[1 +  f_{\mathrm{DL}}\left(\frac{R_{2}}{R_{1}}\right)\right]\sin\eta.
\end{align}
\end{subequations}
Eq.~\eqref{eq:SkyrmionVelocity} represents the main results of this manuscript:
(i)~Topologically trivial structures can experience a skyrmion Hall effect: 
Although Eq.~\eqref{eq:SkyrmionVelocity} is independent of the topological charge, the magnetic structure does not move along the SOT spin velocity ($\vect{v}^d \nparallel \vect{v}^{\mathrm{eff}}$). 
(ii)~In the limit of vanishing field like torques or for $R_1=R_2$, within the rigid particle ansatz the skyrmion Hall angle is independent of the strength of the SOTs as well as of the specific radial profile and shape, it does, however, depend on the helicity degree of freedom. 
In the limiting cases of a pure N\'eel (Bloch) type, the antiferromagnetic skyrmion moves along (perpendicular) to $\vect v^{\mathrm{eff}}$.

\begin{figure}
	\centering
	\includegraphics[scale=0.52]{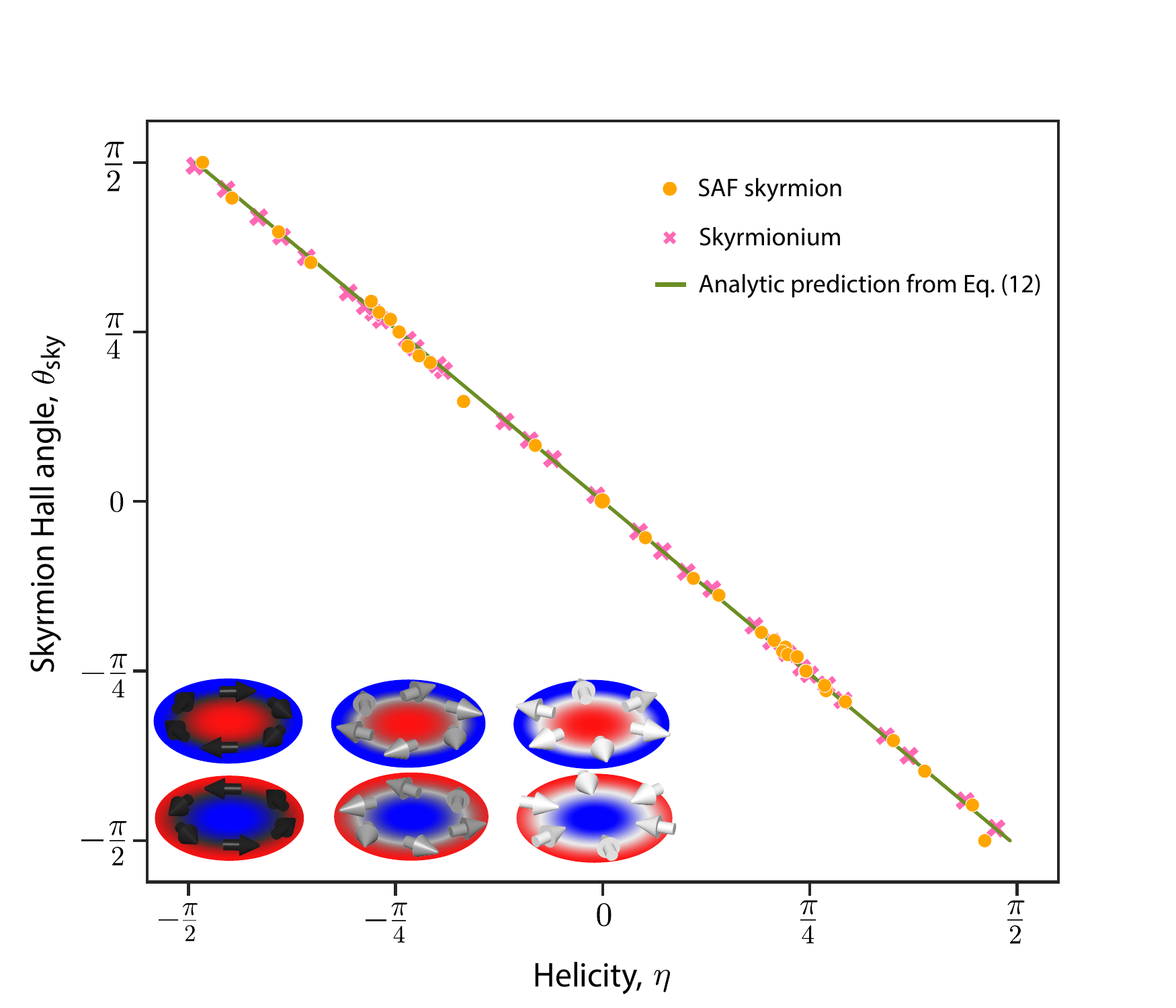}
	\caption{Dependence of the skyrmion Hall angle $\theta_\text{Sky}$ on the helicity degree of freedom of topologically trivial skyrmionic structures.
	}	
	\label{fig:helicity_hall}
\end{figure}

For the simplest case $R_{1} = R_{2}$, the skyrmion Hall angle $\theta_{\textrm{sky}} = \arctan(v_\perp / v_\parallel) $, i.e.\ the angle between $\vect v^{\mathrm{eff}}$ and $\vect v^d$, is given by 
\begin{equation}
	\label{eq:Hallangle}
	\tan\theta_{\textrm{sky}} = - \tan\eta.
\end{equation}
These analytical predictions have been confirmed for skyrmions in SAFs and skyrmioniums by means of micromagnetic simulations using \texttt{MuMax}\textsuperscript{3}~\cite{Vansteenkiste2014}. 
 In Fig.~\ref{fig:helicity_hall} we show the Skyrmion Hall angle as a function of helicity for various systems in the limit where Eq.~\eqref{eq:Hallangle} applies. For details of the numerics see App.~\ref{sec:Methods}.

\section{Discussion and Conclusion}

As a central result we find that topologically trivial magnetic structures can obey a skyrmion Hall effect when driven by SOTs. The helicity of the topologically trivial skyrmionic-like structures is crucial for the direction of motion of their SOT driven dynamics. This contradicts the usual understanding that associates the skyrmion Hall angle just to the topological charge, as is the case for STT driven dynamics (see App.~\ref{app:STT}).

Our work motivates a re-evaluation of the experimental results obtained for skyrmions in SAFs~\cite{Dohi2019}. In this experiment, the authors claimed to have observed only a non-significant skyrmion Hall angle for synthetic antiferromagnetic skyrmions compared to ferromagnetic skyrmions. The larger skyrmion Hall effect in the ferromagnetic samples, may however, be originated in pinning~\cite{Mueller2015, Litzius2020}.  
In particular the role of impurities in synthetic antiferromagnets needs to be clarified in future studies concerning such topologically trivial composite skyrmion structures.
Also interesting for future studies is  the potential influence of the helicity of the corresponding electronic Hall effect.

We emphasize that the predicted skyrmion Hall angle for topologically trivial magnetic structures is independent of the microscopic details and the different physical mechanisms that determine the helicity.  The strongest mechanism is typically associated to chiral interactions such as DMI~\cite{Kim2018, McKeever2018}.
Weaker influences on the helicity are, for example, given by dipolar fields~\cite{Knoester2014}, which in antiferromagnets are typically much smaller compared to ferromagnets. Another source of intermediate helicity are thermal excitations of skyrmions~\cite{Litzius2020} or more generally any type of excitations~\cite{Sun2018, Kravchuk2019, Vakili2020}. In particular for the last reason, we point out that tunning the skyrmion Hall angle to zero will always require  fine tuning.
Furthermore, for skyrmion-like structures stabilized in frustrated magnets, the helicity degree of freedom is a Goldstone mode and can be manipulated by electrical fields~\cite{Zhang2018,Xia2020}.
An advantage of the skyrmion Hall angle dependence on the helicity is the possibility to control the motion direction of skyrmionic structures by changing the helicity. This can, for example, be done with electrical currents and voltage controlled DMI~\cite{Srivastava2018}.

\section{Acknowledgements}
We thank Takaaki Dohi for fruitful discussions and Jonas Nothhelfer for support on the numerical simulations.
We acknowledge funding from the Deutsche Forschungsgemeinschaft (DFG, German Research Foundation) from projects No. 320163632 (Emmy Noether), No. 403233384 (SPP Skyrmionics) and TRR 173 – 268565370 Spin + X: spin in its collective environment (project B12).
J.L. was  supported  by  the  Fonds  Wetenschappelijk Onderzoek (FWO-Vlaanderen) with (senior)  postdoctoral research  fellowships.
Furthermore, this research was supported in part by the National Science Foundation under Grant No. NSF PHY-1748958.

\begin{appendix}

\section{Thiele equation for skyrmioniums and antiferromagnetic skyrmions}
\label{app:Thiele}

In general, the Thiele equation for a rigidly moving magnetic structure, i.e.\ $\vect{m}(\vect{r},t) = \vect{m}(\vect{r} - \vect v_{d}t)$, in the $xy$ plane is obtained by projecting the LLG Eq.~\eqref{eq:LLG} onto $\vect{m}\times\partial_{a}\vect{m}$ where $a = x,y$.

In the skyrmionium case, we consider a magnetic configuration $\vect{m} = \vect{m}_{1} + \vect{m}_{2}$, 
where $\vect{m}_{1}$ and $\vect{m}_{2}$ are the two concentric skyrmionic profiles with different radii. Therefore it is $\partial_{a}\vect{m} = \partial_{a}\vect{m}_{1} + \partial_{a}\vect{m}_{2}$. 
We also assume that $\partial_{a}\vect{m}_{1}\cdot\partial_{b}\vect{m}_{2} \approx 0 $ for $a \neq b$. Therefore, if we project Eq.~\eqref{eq:LLG} onto $\vect{m}_{i}\times\partial_{a}\vect{m}_{i}$, we obtain Eq.~\eqref{eq:Thiele} for the skyrmionium.

For the (synthetic) antiferromagnetic skyrmion, we consider that the total magnetization profile is represented by two skyrmions placed at different layers/sublattices, which are coupled antiferromagnetically. The magnetization dynamics is described by a pair of LLG equations,
\begin{equation}\label{eq:LLG_sub}
	\partial_t\vect{m}_{i}=-\gamma\vect{m}_{i}\times\vect{H}_{i}^{\rm eff}+\alpha\vect{m}_{i}\times\partial_t\vect{m}_{i}+\vect{T}(\vect{m}_{i}).
\end{equation}
with $i$ labeling the different layer/sublattice.
By projecting Eq.~\eqref{eq:LLG_sub} onto  $\vect{m}_{i}\times\partial_{a}\vect{m}_{i}$, we obtain Eq.~\eqref{eq:Thiele} for the (synthetic) antiferromagnetic skyrmion.

To obtain explicit expressions for the tensors entering Eq.~\eqref{eq:Thiele}, we assume a radially symmetric skyrmion, which can be described in terms of spherical coordinates as
\begin{multline}
	\label{eq:SkyAnsatz}
	\vect{m_1}(r, \psi) = \cos\theta(r)\hat{\vect{z}}  \\+
	\sin\theta(r)\left(\cos(\psi + \eta)\hat{\vect{x}} + \sin(\psi + \eta)\hat{\vect{y}}\right),
\end{multline}
where  $r$ and $\psi$ are polar coordinates. 
The profile function $\theta(r)$ of the first skyrmion varies from $0$ at $r=0$ to $\pi$ at infinity with $\theta(R_1)=\pi/2$ defining $R_1$ the radius of the first skyrmion.

For (synthetic) antiferromagnetic skyrmions $\vect{m}_2\approx-\vect{m}_1$, i.e., $R_2 = R_1$ and the angular function $\theta(r)$ defining the profile of the skyrmion gets shifted by $\pi$: $\theta_{2}(r) \rightarrow \theta_{1}(r) + \pi$ while the helicity angle $\eta$ of the skyrmion remains the same to obtain $\vect{m}_2$. 

For the skyrmionium, the outer skyrmion has a different radius $R_2$ and potentially a different profile than the inner skyrmion. We approximate 
the profile of the outer skyrmion $\theta_{2}(r - R_{2})\approx \theta_{1}(r - R_{1}) + \pi$, i.e., and inverted and outward shifted profile function. 

Note that the mathematical form of the ansatz in Eq.~\eqref{eq:SkyAnsatz} is independent of specific material parameters. All material parameters and specific interactions are included in the profile function $\theta(r)$ and the value of the helicity angle $\eta$. 

Plugging the ansatz \eqref{eq:SkyAnsatz} into Eqs.~\eqref{eq:Q}-\eqref{eq:ThieleSOTs} yields a topological charge of modulus one, but an opposite sign for each skyrmion, i.e.\ $Q_{1} = - Q_{2} $, as expected. Thus, the Magnus force term proportional to $\vect G_i$ acts in opposite directions for the skyrmions with different polarities, $\vect G_1 = - \vect G_2$, and their sum, i.e., the total Magnus force on the magnetic quasiparticle cancels.
The result of the dissipative tensor is given by, 
\begin{subequations}\label{eq:Dtensorsk}
	\begin{align}
		(\vect{D}_{i})_{xx} &= (\vect{D}_{i})_{yy} \equiv D_{i},\\
		(\vect{D}_{i})_{xy} &= (\vect{D}_{i})_{yx} = 0, 
	\end{align}
\end{subequations}
with $D_{1} = \pi\int dr\,r\left((\partial_{r}\theta(r))^2 + \frac{\sin^2\theta(r)}{r^2}\right)$.
For (synthetic) antiferromagnetic skyrmions the dissipative tensor is independent of the sublattice degree of freedom, thus $D_{1} = D_{2} = D$, as expected from Eq.~\eqref{eq:Dtensor}. 
For the SOT terms we obtain Eqs.~\eqref{eq:SOTansatz} with $\tau_{\rm FL }  = \pi\int dr r \left(\sin\theta_1 + r \cos\theta_1\partial_{r}\theta_1\right)$ and $\tau_{\rm DL} = \pi\int dr r \left(\cos\theta_1 \sin\theta_1 + r \partial_{r}\theta_1\right)$.

\section{Skyrmion Hall angle for spin-transfer torque driven skyrmionic structures}
\label{app:STT}

Spin transfer torques (STTs) can also produce a motion of the skyrmion. They are given by \cite{Li2004,Thiaville2005}
\begin{equation}
	\vect{T}^\text{STT}(\vect{m}) = (\vect{v}^\text{eff}_{s}\cdot\nabla)\vect{m} + \beta \vect{m}\times (\vect{v}^\text{eff}_{s}\cdot\nabla)\vect{m},
\end{equation}
where $\vect{v}^\text{eff}_{s} = \mu_{B} P \vect{j} /(\gamma e M_{s}^2)$ is the effective spin velocity with $\mu_{B}$ being the Bohr magneton, 
$P$ is the current polarization rate, and $\vect{j}$ is the applied current density~\cite{Tomasello2014}.

\begin{figure}[t]
\centering
\includegraphics[width=\linewidth]{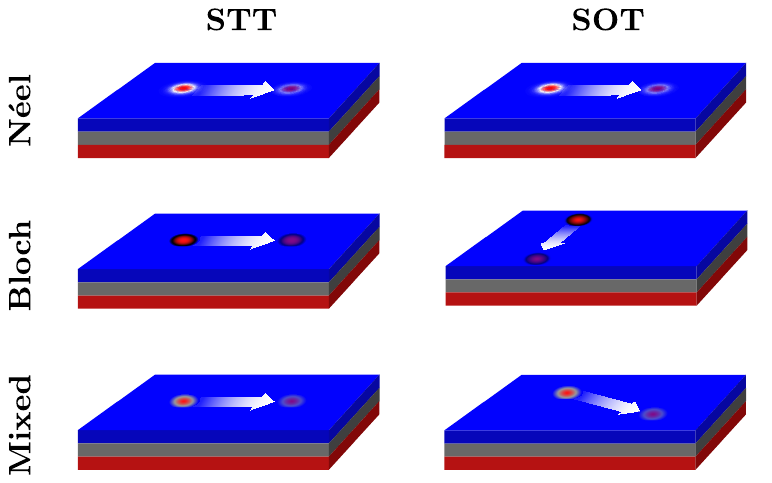}
\caption{Direction of skyrmion motion in a SAF subject to STTs (left column) or SOTs (right column), for N\'eel, Bloch and mixed-type skyrmions. For a spin current applied from left to right, an STT driven skyrmion moves always along the current direction, while it generally moves at an angle to it for the SOT. Color coding is as presented in Fig. \ref{fig:Forces}(b).}
\label{fig:SAFdynamics}
\end{figure}

The Thiele equation for a STT-driven rigid skyrmionic textures (i.e.\ the analog of Eq.~\eqref{eq:Thielesum}) is given by~\cite{Everschor2011}
\begin{equation}\label{eq:ThieleSTT}
	\sum_{i=1,2} - \vect{G}_i \times  (\vect{v}^d - \vect{v}^\text{eff}_{s})- \vect{D}_{i}(\alpha  \vect{v}^d - \beta \vect{v}^\text{eff}_{s})+ \gamma\vect{F}^{\rm int}_{i} = \vect{0}.
\end{equation}
For an antiferromagnetic skyrmion and a skyrmionium the gyrocoupling tensors $\vect{G}_{1} = -\vect{G}_{2}$ and the mutual interaction force $\vect{F}^{\rm int}_{1}= -\vect{F}^{\rm int}_{2}$  are antisymmetric, while the viscosity tensor $\vect{D}_{1} \neq - \vect{D}_{2}$ is not antisymmetric. With this Eq.~\eqref{eq:ThieleSTT} reduces to
\begin{equation}
	\vect{v}_{d} = \frac{\beta}{\alpha}\vect{v}_{s}^{\mathrm{eff}}.
\end{equation}
Thus, an STT-driven topologically trivial magnetic structure moves along the spin current independent of its helicitiy, i.e., the skyrmion Hall angle for STT driven topologically trivial magnetic structures vanishes, see Fig.~\ref{fig:SAFdynamics}.~\cite{Barker2016,Salimath2020}

\section{Micromagnetic Simulation details}\label{sec:Methods}
All simulations are performed using the micromagnetic code MuMax$^3$\cite{Mulkers2017}. The code was extended to allow for the simulation of a system under the influence of both, interfacial and bulk, DMIs.
For the discretization we used a cubic cell size with an edge length of 1nm. 
The ferromagnetic system for the skyrmionium as well as the individual magnetic layers of the synthetic antiferromagnet are described by the energy functional 
\begin{equation}
  \begin{aligned}
  E[\vect{m}] =& \int d^3r \big[A^{\rm ex}(\nabla \cdot \vect{m})^2 - K_z(\vect{m}\cdot \hat{z})^2\big] \\
&+ E^\text{DD} + E^{\rm IDMI} + E^{\rm BDMI},
  \end{aligned}
  \label{eq:energy}
\end{equation}
where $A^\text{ex}$ is the (interlayer) exchange stiffness, and $K_z>0$ is the uniaxial anisotropy strength. 
The term $E^\text{DD} = -\frac{1}{2}\mu_0M_s\vect{m}_i\cdot\vect{H}_{m,i}$ accounts for the dipole–dipole
interaction, with $\mu_0$ being the vacuum permeability, and $\vect{H}_{m,i}$ the magnetostatic dipolar field. DMI interactions are introduced to the energy functional through the terms $E_i^{\rm IDMI}[\vect{m}]=\int d^3r D_{\rm int}\vect{m}_i\cdot(\hat{z}\times\nabla)\times\vect{m}_i$, and $E_i^{\rm BDMI}[\vect{m}]=\int d^3r~D_{\rm bulk}\vect{m}_i\cdot(\nabla\times\vect{m}_i)$, denoting interfacial and bulk DMI respectively.
All simulation results shown share the parameter values $M_{s} = 0.58$ MA/m, $A^\text{ex} = 30$ pJ/m 
and $K_z=0.8$ MJm$^{-3}$.
We chose $\xi=-0.02$ and a damping of $\alpha = 0.1$.
Please note that the Skyrmion Hall angle is independent of $\alpha$, as it only appears as a global prefactor in Eq.~\eqref{eq:SkyrmionVelocity}.

\subsection{Skyrmionium Simulations}
For the skyrmionium simulations we considered a sample of size 512$\times$256$\times$1 with periodic boundary conditions. 
For the results shown in Fig.~\ref{fig:helicity_hall} in the main text, we varied the DMI strengths in a range of $D_{\rm int} = -4.2 \cdot 10^{-3}$ J/m$^2$ to  $D_{\rm int} =4.2\cdot 10^{-3}$ J/m$^2$ and $D_{\rm bulk} = -4.2\cdot 10^{-3}$ J/m$^2$  
to $D_{\rm bulk} = 4.2\cdot 10^{-3}$ J/m$^2$ to produce stable skyrmioniums with various helicities.

\subsection{Simulations in Synthetic Antiferromagnets}
\begin{figure}[t]
	\centering
	\includegraphics[width=\linewidth]{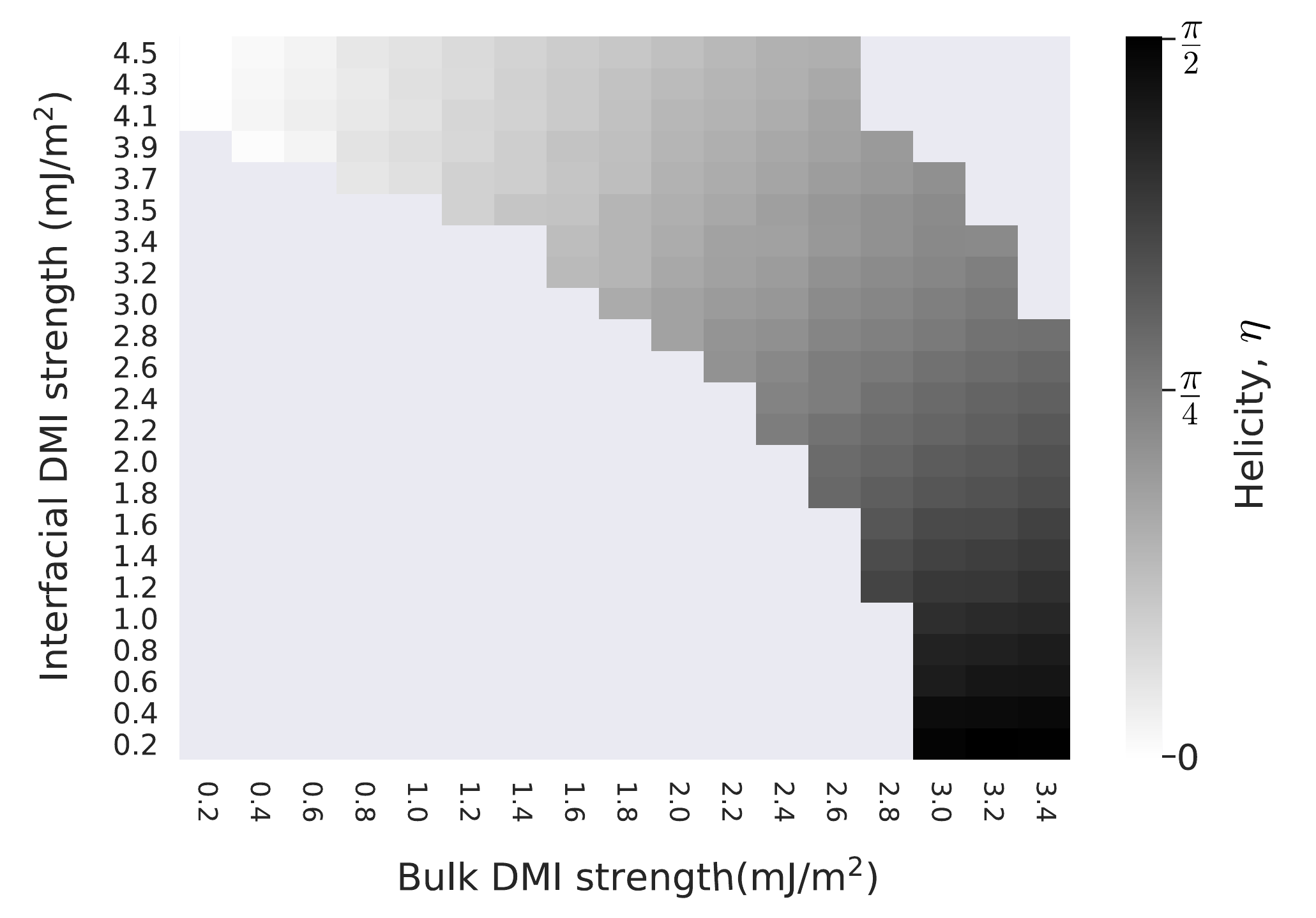}
	\caption{Helicity of the synthetic antiferromagnetic skyrmion for different bulk D$_\text{bulk}$ and interfacial DMI strengths D$_\text{int}$.
}
	\label{fig:heat_hall}
\end{figure}

For the SAF simulations we modelled a three layer system of dimensions 256nm$\times$256nm with periodic boundary conditions, where the height of each layer is chosen to be 1nm, see Fig.~\ref{fig:SAFdynamics} for a sketch. The top and bottom layer are coupled antiferromagnetically with strength $A$ via the additional term $E^{\rm AFM}=-\int d^3r A(\vect{m}_{1}\cdot\vect{m}_{2})$ such that the energy functional for each sublattice $E_i$ becomes 
\begin{equation}
  \begin{aligned}
  E_i[\vect{m}] =& \int d^3r \big[A^\text{ex}(\nabla \cdot \vect{m}_i)^2  - K_z(\vect{m}_i\cdot \hat{z})^2 + A(\vect{m}_{1}\cdot\vect{m}_{2})\big] \\
&+ E_i^\text{DD} + E_i^{\rm IDMI} + E_i^{\rm BDMI},
  \end{aligned}
  \label{eq:SAFenergy}
\end{equation}
where $i = \{1, 2\}$ is the layer index (1 for the top, 2 for the bottom).
For the results shown in Fig.~\ref{fig:helicity_hall} in the main text, we used $A = 5.0\times10^{-13}$ J/m and varied the DMI parameters in the ranges $D_{\rm int} = 0 - 4.5$ mJ/m$^2$ and $D_{\rm bulk} = - 3.4 - 3.4$ mJ/m$^2$, yielding the different helicities, as shown in Fig.~\ref{fig:heat_hall}.
We simulated the STT- and SOT-driven dynamics of synthetic antiferromagnetic skyrmions with various helicities, where Fig.~\ref{fig:SAFdynamics} summarizes and confirms the key results obtained in the main part and App.~\ref{app:STT}. 

\end{appendix}

\vspace{5mm}
\section{Data availability}
All relevant data presented in the manuscript and in the Supplementary
Information supporting the findings of this study are available from the
corresponding authors upon reasonable request.

\end{document}